\documentclass[12pt]{iopart}

\usepackage{xcolor}%
\usepackage{natbib}
\usepackage{graphicx}%
\usepackage{amsmath,amssymb,amsfonts}%
\usepackage{algorithm}
\usepackage{algorithmic}
\begin{document}

\title{One-shot omnidirectional pressure integration through matrix inversion}

\author{Fernando Zigunov$^{1,2}$ \& John J. Charonko$^1$}

\address{$^1$ Los Alamos National Laboratory, Physics Division, Los Alamos, USA}
\address{$^2$ Syracuse University, Department of Mechanical and Aerospace Engineering, Syracuse, USA}
\ead{fzigunov@syr.edu}
\ead{john.charonko@lanl.gov}
\vspace{10pt}
\begin{indented}
	\item[]January 2024
\end{indented}

\begin{abstract}
	In this work, we present a method to perform 2D and 3D omnidirectional pressure integration from velocity measurements with a single-iteration matrix inversion approach. This work builds upon our previous work, where the rotating parallel ray approach was extended to the limit of infinite rays by taking continuous projection integrals of the ray paths and recasting the problem as an iterative matrix inversion problem. This iterative matrix equation is now ``fast-forwarded'' to the ``infinity'' iteration, leading to a different matrix equation that can be solved in a single iteration, thereby presenting the same computational complexity as the Poisson equation. We observe computational speedups of $\sim10^6$ when compared to brute-force omnidirectional integration methods, enabling the treatment of grids of $\sim 10^9$ points and potentially even larger in a desktop setup at the time of publication. Further examination of the boundary conditions of our one-shot method shows that omnidirectional pressure integration implements a new type of boundary condition, which treats the boundary points as interior points to the extent that information is available. Finally, we show how the method can be extended from the regular grids typical of particle image velocimetry to the unstructured meshes characteristic of particle tracking velocimetry data.
	
\end{abstract}

\section{Introduction}

The reconstruction of mean and instantaneous pressure fields from particle image velocimetry (PIV) measurements can offer significant insight on flow physics. In time-resolved flow fields, the knowledge of instantaneous pressure can enable the reconstruction of pressure-velocity correlations, which are of interest to the improvement of Reynolds-Averaged Navier-Stokes (RANS) equation models. For averaged flow fields, the knowledge of the pressure field can offer the opportunity to better understand sources of aerodynamic drag and, in compressible high-speed flows, the knowledge of the average pressure and velocity fields enables the inference of the temperature and density fields through a few assumptions \citep{vanOudheusden2008}, which can greatly improve our ability to assess accuracy of CFD computations and further our understanding of three-dimensional compressibility phenomena. Therefore, the computation of the pressure field from PIV has been explored in depth throughout the last decades.

The presence of errors and spatial noise in the velocity fields obtained from PIV, however, poses a challenge to performing integration of the pressure fields. \citet{charonko_assessment_2010} noted that even in simple flows, such as a 2D Taylor vortex, solving the pressure-poisson equation results in very large errors for small introduced errors in the underlying velocity fields. Furthermore, the pressure-poisson equation requires boundary conditions on all surfaces of the domain, which seems to be a major source of error when Neumann boundary conditions are specified, as detailed by \citet{liu_error_2020} and further explained by \citet{pan_error_2016}. The omnidirectional ray integration methods proposed over the last two decades (rectangular virtual boundary, \citep{liu_measurements_2003}, circular virtual boundary \citep{Liu2006}, rotating paralllel ray \citep{liu_instantaneous_2016}) perform significantly better than the pressure-poisson solver, as found by \citet{charonko_assessment_2010}; and require no specification of boundary conditions because they iteratively converge the boundary values to satisfy the pressure gradient field specified in the interior points by the PIV velocity field. This is especially useful for experiments where parts of the boundary still contain regions with strong variation in pressure due to limitations in the field of view. 

The computational cost to compute all the ray integration paths in omnidirectional methods, however, is very large and prohibitive for 3D domains. \citet{wang_gpu-based_2019} implemented a GPU-accelerated 3D omnidirectional integrator (ODI), achieving moderate integration times of a few minutes for small 3D grids ($100\times 47\times 38 = 1.78\times 10^5$ points). We recently proposed \citep{Zigunov2023mst} a different approach to build an ODI solver, based on converting the ODI ray integration process into a matrix inversion approach that updates the boundary (and interior) nodes of the domain in a similar fashion as the original ODI methods. We refer to this approach as the \textit{``Iterative Matrix ODI''} (I-MODI), and significant speedups of about 3 orders of magnitude were possible in a GPU-accelerated implementation when compared to the work of \citet{wang_gpu-based_2019}.

During the development of the I-MODI method, we found that the number of iterations required to converge the boundaries with the iterative update equation can reach hundreds or even thousands when the domain boundaries contain significant pressure variations, such as when the domain includes a wall. We further explore the MODI approach to present a generalization of the method that requires only one matrix inversion to reach the converged state of the previously proposed I-MODI method. We will refer to this new method as the \textit{``One-Shot Matrix ODI''} (OS-MODI). This approach enables the robust computation of pressure from PIV for very large 2D and 3D domains, of $10^6$ to almost $10^9$ points, in a few seconds to a few minutes. We further extend the discussion by pointing out the nature of the boundary conditions that the omnidirectional ray integration approach implements, which appears to treat the boundary points like interior points to the extent information is available.

\section{Generalization of Matrix-based Omnidirectional Integration} \label{MatrixOmni}

\subsection{Method Review}  \label{sec:Concept}

In this section, a brief review of the iterative matrix-based omnidirectional pressure integration method described in our previous work \citep{Zigunov2023mst} is presented for context. 

We start with the parallel-ray omnidirectional integration approach \citep{liu_instantaneous_2016}, where multiple rays are cast from outside of the integration domain, entering through a boundary and exiting through the opposite boundary of the domain. The parallel-ray approach considers a large number of lateral shifts for a given ray angle, and a large number of angles; in nested \textit{for} loops that can be parallelized on a GPU \citep{wang_gpu-based_2019}. The pressure, being a scalar field, is independent of integration path and therefore the pressure at any point can be found by averaging the very large number of rays considered. For a given ray at a given point $P_C$, the pressure is computed by considering an adjacent point $P_a$:

\begin{equation}
	P_C = P_a + \int ^ {\vec x_C} _{\vec x_a} \frac{\partial P}{\partial \vec r} \cdot \vec r
\end{equation}

\noindent where $\vec{r}$ is the ray direction and $\partial P/\partial \vec{r}$ is the pressure gradient field, which can be found by isolating the velocity field in the momentum equation (see \citet{Zigunov2023mst} for details).

Let us consider the following two assumptions for the following discussion: (1) The grid is uniformly-spaced and consists of rectangular or box-like elements and (2) The ray paths are discretized such that they always cross the center of the cells; we can focus our attention to an individual cell of the grid. 

\begin{figure*} [h]
	\centering
	\includegraphics[width=100mm]{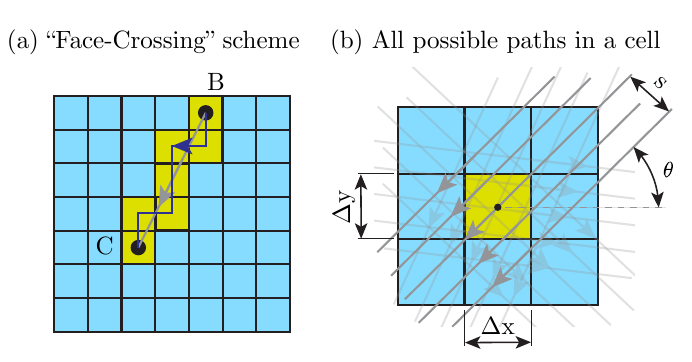}
	\caption{\centering Graphical description of the integration scheme discussed \citep{Zigunov2023mst}.}
	\label{fig:line-integration}
\end{figure*}

A representation of the discretization of the ray path is shown in Figure \ref{fig:line-integration} (a). Considering the spacing between the rays $\Delta s$ and the angular step $\Delta \theta$ in the parallel-ray omnidirectional method, we can take the limit where $\Delta s\rightarrow 0$ and $\Delta\theta\rightarrow 0$ (i.e., taking the number of rays to infinity). This process of considering all possible rays is represented graphically in Figure \ref{fig:line-integration} (b). Doing so enables the definition of the following iterative update equation for the pressure at an arbitrary point $C$ with neighboring points $j$:

\begin{equation} \label{eq:ExplicitEqn_FaceCrossing}
	P_C^{n+1}= \sum_i \sum_j w_j^i P_j^i - \sum_{j} \Delta j (w_{j}^{n+1} + w_{j}^{n}) f_j(j)
\end{equation}

\noindent where the iteration $i$ can take the values $i=\{n,n+1\}$ (i.e., previous and next iteration), and the cell $j$ can take the values $j=\{C,E,W,N,S,F,B\}$, representing the cardinal directions Center, East, West, North, South, Front and Back, respectively (for ``face crossing scheme'' described in \citet{Zigunov2023mst}). 

The index $j$ also indicates the direction, so if $j=E$ we use $\Delta j=\Delta x$ and $f_j = f_x$, and if $j=W$ we use $\Delta j=-\Delta x$ and $f_j = f_x$; and similarly for the other two directions. The source function $f_j$ (at direction $j$) in the last summation is the value of the pressure gradient $\partial P/\partial \vec{r}$ at the face $j$ in direction $j$, which can be found with an appropriate discretization scheme.

The weights $w_i^j$ in Equation \ref{eq:ExplicitEqn_FaceCrossing} represent the fraction of the total rays coming from a given adjacent cell $j$. If cell $j$ is near a boundary of the domain, some rays will start at cell $j$, requiring the value of the cell at the $n^{th}$ (previous) iteration. The rays that start somewhere else and simply go through cell $j$ use the value of the cell at the $(n+1)^{th}$ (next) iteration. These weights can be computed using integrals based on the geometry of the problem, and a thorough treatment of these integrals is presented in our previous publication \citep{Zigunov2023mst} for grids composed of 2D rectangular elements or 3D box-like elements. 

Equation \ref{eq:ExplicitEqn_FaceCrossing} can be solved iteratively with an implicit matrix solver, and it is shown in \citep{Zigunov2023mst} that a three orders of magnitude speedup in the computational time can be achieved when compared to the parallel-ray omnidirectional method, along with significant improvement in resilience to noise in the underlying velocity fields. Equation \ref{eq:ExplicitEqn_FaceCrossing} will be called through this manuscript the Iterative Matrix OmniDirectional Integrator (I-MODI). However, the iterative form of Equation \ref{eq:ExplicitEqn_FaceCrossing} still poses a challenge because hundreds of iterations are required to reach convergence for realistic pressure fields involving missing regions or walls with significant pressure variations at a boundary.

\subsection{Solving for Pressure in a Single Shot} \label{sec:OneShot}

Upon closer inspection of Equation \ref{eq:ExplicitEqn_FaceCrossing}, one can ask the following question: Is an iterative update of the pressure field really required? Or can we ``fast-forward'' to the last iteration ($n\rightarrow \infty$)?

If we assume the method converges, at the ``infinity iteration'' $P_j^n=P_j^{n+1}=P_j$ and we can combine the terms in Equation \ref{eq:ExplicitEqn_FaceCrossing}:

\begin{equation} \label{eq:SingleShot1}
	(1-w_C^n)P_C - \sum_j (w_j^n + w_j^{n+1}) P_j = -\sum_{j} \Delta j (w_{j}^{n+1} + w_{j}^{n}) f_j(j)
\end{equation}

The combination of the weights $(w_j^n + w_j^{n+1})$ in Equation \ref{eq:SingleShot1} is key for the simplification of this problem. Since the $n^{th}$ and $(n+1)^{th}$ iteration weights are combined, we now only need to consider the fraction of rays coming from cell $j$ \textit{regardless of whether they were previously a boundary or not}. Therefore, we now only need to assess if the cell $j$ is itself a boundary. Following our previous work, we define a boolean value $b_j$ for each cell, defining whether there is missing data (NaN) in the corresponding cell for the source field $\vec{f}$:

\begin{equation}
	b_j=\begin{cases}
		0 \;\; \textrm{if } \vec{f}(j) \neq \textrm{NaN} \\
		1 \;\; \textrm{if } \vec{f}(j) = \textrm{NaN}
	\end{cases}
\end{equation}

This enables working with arbitrary boundary topologies that may have missing data due to laser reflections, lack of illumination, etc; which is a large majority of cases in realistic PIV vector fields. We also define $\tilde{b}_j=1-b_j$ as the boolean NOT operation on $b_j$ to define where data is available instead. 

In two-dimensions for the ``face-crossing'' scheme, the sum of weights $(w_{j}^{n} + w_{j}^{n+1})$ is:

\begin{equation}
	(w_{j}^{n} + w_{j}^{n+1})=\begin{cases}
		\tilde{b}_j\;\frac{\Delta y}{2(\Delta x + \Delta y)} \;\; \textrm{if } j=\{E, W\} \\
		\tilde{b}_j\;\frac{\Delta x}{2(\Delta x + \Delta y)} \;\; \textrm{if } j=\{N, S\}
	\end{cases}
\;\; \textrm{(2D, face-crossing)}
\end{equation}

For 3D grids, the expression takes a similar form:

\begin{equation}
	(w_{j}^{n} + w_{j}^{n+1})=\begin{cases}
		\tilde{b}_j\;\frac{\Delta y\Delta z}{2(\Delta x\Delta y + \Delta x\Delta z+\Delta y \Delta z)} \;\; \textrm{if } j=\{E, W\} \\
		\tilde{b}_j\;\frac{\Delta x\Delta z}{2(\Delta x\Delta y + \Delta x\Delta z+\Delta y \Delta z)} \;\; \textrm{if } j=\{N, S\} \\
		\tilde{b}_j\;\frac{\Delta y\Delta z}{2(\Delta x\Delta y + \Delta x\Delta z+\Delta y \Delta z)} \;\; \textrm{if } j=\{F, B\}
	\end{cases}
\;\; \textrm{(3D, face-crossing)}
\end{equation}

The pattern is evident for both cases, and is simply the ratio between the corresponding face area $A_j$ and the total surface area of the parallelepiped $A_{tot}$:

\begin{equation} \label{eq:Otherweights}
	(w_{j}^{n} + w_{j}^{n+1})=\tilde{b}_j \frac{A_j}{A_{tot}}
\end{equation}

This fact is generalizable for any convex cell shape, which is explored further in Section \ref{sec:ArbGrids}. 

We note that the center cell weight $w_C^n$ in Equation \ref{eq:SingleShot1} is related to all rays that initiate at cell $C$ because the ray direction is such that the first cell of the domain is cell $C$. As it was shown in \citep{Zigunov2023mst} for the parallelepiped grid, by exclusion $w_C^n$ must be:

\begin{equation} \label{eq:Centerweight}
	w_C^n=\sum_j b_j \frac{A_j}{A_{tot}}
\end{equation}

Replacing Equations \ref{eq:Otherweights} and \ref{eq:Centerweight} in Equation \ref{eq:SingleShot1} and realizing that summing all boundary and non-boundary faces gives the total cell surface area, i.e. $\sum_j (b_j A_j + \tilde{b}_j A_j) = A_{tot}$:

\begin{equation} \label{eq:FaceCrossing2}
	P_C	\sum_j  \frac{\tilde{b}_j A_j}{A_{tot}} - \sum_j  \frac{\tilde{b}_j A_j}{A_{tot}} P_j = -\sum_{j} \Delta j  \frac{\tilde{b}_j A_j}{A_{tot}} f_j(j)
\end{equation}

If we now define the ``interior area'' $A_{int}$ as the area of the faces that are \textit{not domain boundaries}:

\begin{equation}
	A_{int} = \sum_j  \tilde{b}_j A_j
\end{equation}

\noindent then we can further simplify Equation \ref{eq:FaceCrossing2}:

\begin{equation} \label{eq:FaceCrossing3}
	P_C = \sum_j  \frac{\tilde{b}_j A_j}{A_{int}} \big( P_j - f_j(j) \Delta j  \big)
\end{equation}

Equation \ref{eq:FaceCrossing3} can be written as a matrix form for 2D rectangular grids or 3D parallelepiped grids. Solving this equation enables reaching the same result as the converged omnidirectional integration through a single iteration, and is therefore called the ``One-Shot'' Matrix OmniDirectional Integration (OS-MODI) method.

Note Equation \ref{eq:FaceCrossing3} exposes the weights as simple area ratios ($A_j/A_{int}$), which can be computed on-the-fly with minimal cost. This improves the memory usage of the OS-MODI solver, as it eliminates the need to store the weight matrices. The I-MODI method requires $16\cdot 8 (N_x\times N_y\times N_z)$ bytes of GPU memory to store 16 scalar fields required (7 for the CG solver, 6 for the weight matrices, 3 for the source field $x,y,z$ components). The non-iterative nature of the OS-MODI method means we only require 6 fields for the CG solver, plus 3 fields for the source components. As the weight functions are so simple, they do not need to be precomputed, meaning the 6 weight matrices are not required to be stored in memory. This means the OS-MODI solver only requires a total of 9 fields to be stored in memory and $9\cdot 8 (N_x\times N_y\times N_z)$ bytes of GPU memory when computing with double precision and $9\cdot 4 (N_x\times N_y\times N_z)$ bytes at single precision. For the 16GB memory of the Nvidia RTX A4000 GPU used in this study, this means a maximum of 477 million grid points can be fitted on memory at single-precision, enabling a grid of up to $780^3$ points to be handled.

\section{Relationship to the Pressure-Poisson Equations} \label{sec:Poisson}

If we discretize the Pressure-Poisson equation using a finite volume method on 2D rectangular cells and second-order accurate derivatives, we get the following form:

\begin{equation} \label{eq:PressurePoissonDiscrete2D}
	\begin{split}
		\Delta y^2 (P_E - 2P_C +P_W) + \Delta x^2 (P_N - 2P_C +P_S) = \dots \\
		\frac{\Delta x \Delta y^2}{2} [f_x(E)-f_x(W)] + \frac{\Delta x^2 \Delta y}{2} [f_y(N)-f_y(S)] \\
		\textrm{(Pressure-Poisson, interior point)}
	\end{split}
\end{equation}

We can also examine the equation for an interior cell of the iterative omnidirectional integration method (I-MODI) in a 2D rectangular grid, following Equation \ref{eq:ExplicitEqn_FaceCrossing} and yielding:

\begin{equation} \label{eq:OmniDiscrete2D}
	\begin{split}
		\Delta y (P_E - 2P_C +P_W) + \Delta x (P_N - 2P_C +P_S) = \dots \\
		\frac{\Delta x \Delta y}{2} [f_x(E)-f_x(W)] + \frac{\Delta x \Delta y}{2} [f_y(N)-f_y(S)] \\
		\textrm{(I-MODI, interior point)}
	\end{split}
\end{equation}

This equation applies only for interior points. In this case, all values of pressure are coming from the $(n+1)^{th}$ iteration. Note Equations \ref{eq:PressurePoissonDiscrete2D} and \ref{eq:OmniDiscrete2D} have very similar forms, and are the exact same equation when $\Delta x=\Delta y$. The equations are slightly different in the case $\Delta x \neq \Delta y$, owing to the first-order integration performed by the matrix-omnidirectional method being different to the second-order integration performed in the Pressure-Poisson equation. 

%
%
%

To further understand the behavior of the I-MODI method at the boundaries, let's examine the equation implemented for the westmost node in the middle of a 2D isotropic grid (i.e., $\Delta x=\Delta y=\Delta$):

\begin{equation} 
	\begin{split}
		P_C^{n+1} = \frac{1}{4} (P_C^{n} + P_E^{n+1}) + \frac{\sqrt{2}}{8}(P_N^{n+1} + P_S^{n+1}) + \frac{2-\sqrt{2}}{8}(P_N^{n} + P_S^{n}) + \dots \\
		\frac{1}{4} \Delta \bigg[ -\frac{f_x(E)+f_x(C)}{2} - \frac{f_y(N)+f_y(C)}{2} + \frac{f_y(S)+f_y(C)}{2} \bigg]\\ \textrm{(I-MODI, west boundary)}
	\end{split}
\end{equation}

As we move towards the infinity iteration ($n\rightarrow \infty$, ``one-shot method''), the equation at the boundary converges to:

\begin{equation} \label{eq:BC-omni}
	\begin{split}
		P_C = \frac{1}{3} (P_E + P_N + P_S) + \dots \\
		\frac{1}{3} \Delta \bigg[ -\frac{f_x(E)+f_x(C)}{2} - \frac{f_y(N)+f_y(C)}{2} + \frac{f_y(S)+f_y(C)}{2} \bigg]\\ \textrm{(OS-MODI, west boundary)}
	\end{split}
\end{equation}

\noindent where the pressures no longer have superscripts for the ``infinity iteration'', since $P^{\infty+1}=P^\infty$. Note how the form in Equation \ref{eq:BC-omni} is very similar to the interior point equation of the pressure-Poisson equation for the isotropic grid:

\begin{equation} \label{eq:int-poisson}
	\begin{split}
		P_C = \frac{1}{4} (P_E + P_W + P_N + P_S) + \dots \\
		\frac{1}{4} \Delta \bigg[ -\frac{f_x(E)+f_x(C)}{2} + \frac{f_x(W)+f_x(C)}{2} - \frac{f_y(N)+f_y(C)}{2} + \frac{f_y(S)+f_y(C)}{2} \bigg]\\ \textrm{(Poisson, interior point)}
	\end{split}
\end{equation}

In other words, the one-shot matrix omnidirectional (OS-MODI) equation at the boundary is performing a very similar operation to the Poisson kernel, that is, an averaging operation with the surrounding pressure values and source function values. The average is reweighted to consider the missing nodes at the boundary. This is true for all boundaries (i.e., a corner would have a $1/2$ weight, etc.). Therefore, the matrix omnidirectional method is treating the boundary as close as is possible to an interior point considering the missing information.

This is in contrast with the treatment usually used for a Neumann boundary in the Poisson equation. Typically, a Neumann boundary where the forcing function is known would implement a discrete form of the following equation:

\begin{equation} \label{eq:NeumannBC}
	\nabla P \cdot \hat{n} = \vec{f}\cdot \hat{n}
\end{equation}

In the Neumann boundary condition, the component of the pressure gradient normal to the boundary surface is enforced to be the same value as the normal component of the material acceleration forcing term $\vec{f}$. This type of boundary, in a first-order accurate finite difference scheme, implements the following equation at the east face of the westmost node:

\begin{equation} \label{eq:NeumannBCfd}
	\frac{P_C - P_E}{\Delta} = -\frac{f_x(E) + f_x(C)}{2}
\end{equation}

Note that in general, both components (normal and tangential) of the pressure gradient should match the forcing term $\vec{f}$. This is especially the case if the domain is cropped where there still are meaningful variations in the pressure at the boundary, as is often the case in PIV experiments. However, in practice, the vector equation $\nabla P = \vec{f}$ cannot be implemented without adding an extra equation to the discrete version of the Poisson problem, which would make it overdefined. Say, for example, to enforce the tangential component for the westmost node, one would have to either implement:
	
\begin{equation} \label{eq:TangentialBC1}
	\frac{P_N - P_S}{2\Delta} = f_y(C)
\end{equation}

\noindent which is second-order accurate but does not include the pressure at the center node, or alternatively:

\begin{equation} \label{eq:TangentialBC2}
	\frac{P_C - P_N}{\Delta} = -\frac{f_y(N) + f_y(C)}{2}
\end{equation}

\begin{equation} \label{eq:TangentialBC3}
	\frac{P_C - P_S}{\Delta} = \frac{f_y(S) + f_y(C)}{2}
\end{equation}

\noindent which are first-order accurate approximations that now include the center node but add two more equations at the boundary. In other words, it appears that it is impractical to numerically implement a boundary condition for the Poisson equation where $\nabla P = \vec{f}$ is explicitly enforced in both normal and tangential directions, because doing so would over-define the problem by one or two equations per boundary node. When working with exact data, such as in a CFD simulation, the tangential component is automatically enforced. However, if the source term contains errors, then the usage of both normal and tangential components of the source term (as, for example, in ODI methods) should improve the accuracy of the boundary computation.

Curiously, however, Equation \ref{eq:BC-omni} seems to be implementing such a boundary. If one sums Equations \ref{eq:NeumannBCfd}, \ref{eq:TangentialBC2} and  \ref{eq:TangentialBC3} with a ($1/3$) weighting factor for each, the result is exactly Equation \ref{eq:BC-omni}. This suggests that the OS-MODI method, and by extension all ODI methods with a rotating parallel ray approach, implement a boundary condition that attempts to enforce $\nabla P = \vec{f}$ at the boundary for both normal and tangential components in a single equation, with an equal weight to each face (in the isotropic grid case). 

The question left for future research is whether the traditional Neumann boundary condition (i.e., Equation \ref{eq:NeumannBC}) is physically reasonable for PIV pressure reconstruction. In a general material acceleration field that could be cropped in the middle due to a lack of illumination, laser reflections or simply lack of overlap between camera views, it appears that considering both normal and tangential components of the pressure gradient is a more sensible approach, which seems to be implemented naturally by the ODI methods.

\section{One-Shot Method Performance} \label{sec:OSMODI}

In order to establish that the one-shot method proposed in Section \ref{sec:OneShot} provides the same results as the converged iterative method (I-MODI), we analyze the convergence behavior of the iterative method for the $100^3$ grid of the Johns Hopkins University (JHU) turbulence \textit{``isotropic1024coarse''} database \citep{li_turbulence_2008}. Further details on how the data cutout was extracted are provided in our previous publication \citep{Zigunov2023mst}. The $100^3$ cutout is not a decimated version of the original $1024^3$ grid, but a subset of it; therefore being approximately $1/10^{th}$ of the original size in each direction. Both I-MODI and OS-MODI methods were implemented as CUDA-C++ GPU-accelerated code, running on a Nvidia RTX A4000 GPU (retailing $\sim$ \$1,000 at the time of writing).

\begin{figure*} [h]
	\centering
	\includegraphics[width=170mm]{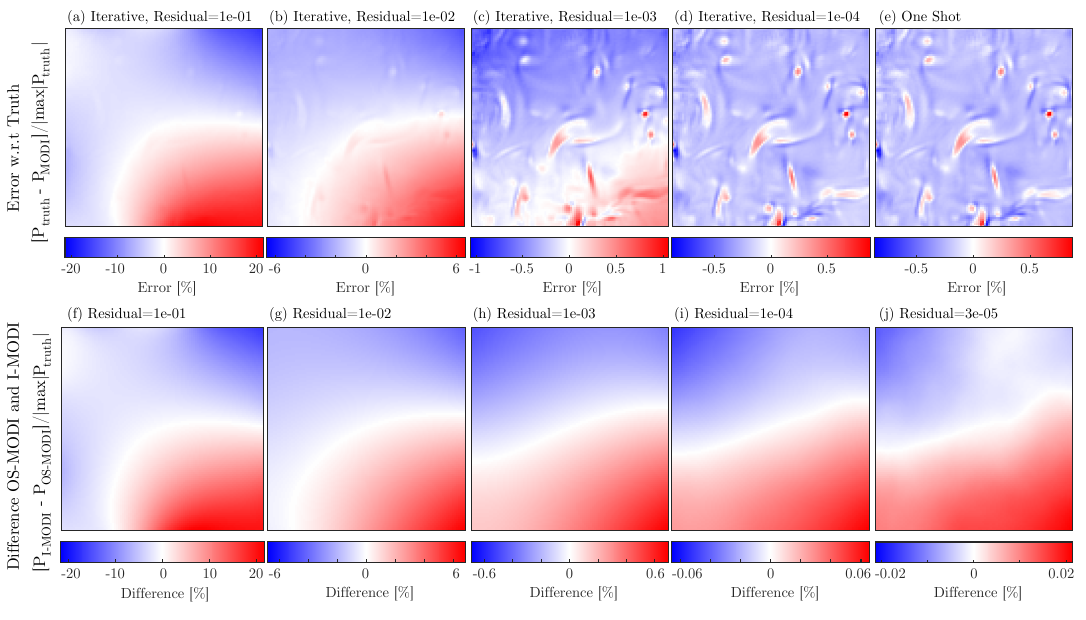}
	\caption{(a-e) Spatial distribution of the error field for the iterative method at various residual levels (a-d) and the one-shot method (e) for the $100^3$ grid for the JHU database. (f-j) Distribution of the difference field between the iterative and one-shot method.}
	\label{fig:convergenceImages}
\end{figure*}

A contour of the error fields is provided in Figure \ref{fig:convergenceImages} for various convergence levels (as defined in \citet{Zigunov2023mst} for the iterative method (I-MODI) in (a-d) and for the one-shot method (OS-MODI) in (e). The error colorbars are variable to encompass the range of errors, which is very large for the higher residuals and becomes very tight (less than $1\%$) for the lower residuals. The one-shot method in Figure \ref{fig:convergenceImages} (e) matches the highest convergence level of the iterative method in Figure \ref{fig:convergenceImages} (d). This trend of the I-MODI method towards matching the OS-MODI as the residual is decreased can also be seen in the second row of images in Figure \ref{fig:convergenceImages} (f-j). As the residual is decreased, note the difference colorbars become tighter and their span is about the order of the residual value, evidencing that full convergence of the omnidirectional pressure solver can be achieved in a single iteration of the solution of the matrix equation with the OS-MODI method. For the residual lower than $10^{-4}$, it is clear that the error is no longer related to the convergence of the iterative method, but potentially to the finite difference approximation of the quantities involved in the calculations.

\begin{figure*} [h]
	\centering
	\includegraphics[width=100mm]{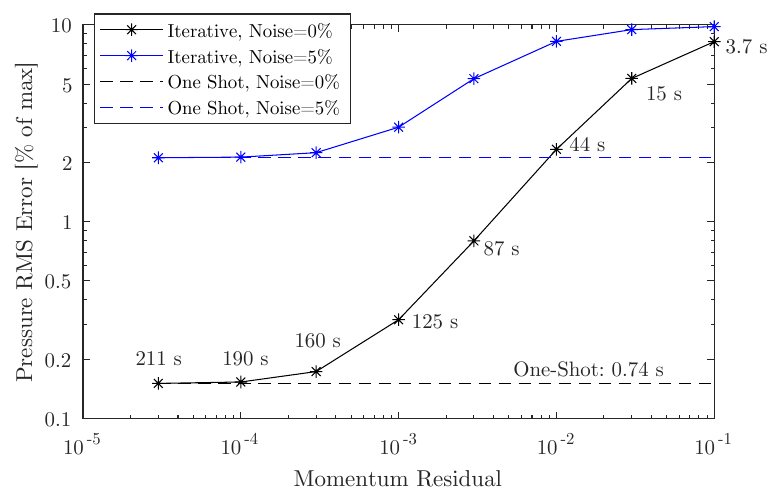}
	\caption{Comparison of RMS value of the error for the I-MODI (``Iterative'') and OS-MODI (``One-Shot'') methods at two noise levels for the $100^3$ grid for the JHU database.}
	\label{fig:convergenceGraph}
\end{figure*}

Further evidence of the convergence behavior is also provided in Figure \ref{fig:convergenceGraph}, where the RMS error is plotted against residual for the iterative method (solid lines) and compared to the one-shot method (dashed lines) for two different noise conditions in the underlying velocity fields. It is evident that the solution achieved by the OS-MODI method is the converged I-MODI method, though the computational time required is significantly lower due to the single-shot nature of the method. For the $100^3$ field considered, a fully converged I-MODI solution requires 211 seconds of computational time in the GPU implementation in both noisy and noiseless cases, whereas the OS-MODI solution only requires 0.74 seconds; meaning we achieve approximately a 285$\times$ speedup. The speedup ratio is problem-dependent because it depends on how many iterations are required for convergence, but it is generally larger for larger grids. Near the largest grid size that can be fitted in this GPU memory ($750^3$ = 421.8 million grid points), the OS-MODI method was solved in 4.7 minutes, demonstrating that even a massive grid with almost half a billion points can be handled within a reasonable time with the proposed approach.

Due to the equivalence of the fully converged I-MODI and the OS-MODI, all the error analysis and behavior for MODI methods presented in \citet{Zigunov2023mst} apply identically to the new OS-MODI method, and thus will not be repeated here.


\section{Arbitrary Grids} 
\subsection{Extending the One-Shot Matrix ODI for Arbitrary Grids} \label{sec:ArbGrids}

For a convex polyhedral cell shape, the surface area illuminated by a parallel set of rays coming from an arbitrary solid angle $(\theta, \phi)$ is exactly the half of the total surface area of the polyhedron $A_{tot}$, except when the rays are parallel to one of the faces. This follows straight from the definition of convexity; i.e., a ray at any angle only crosses the polyhedron surface at exactly two points (unless it lies on an edge/face). 

For a planar face, the area of the projection of the face in a given direction $\hat{r}$ is the unprojected area $A_{face}$ times the dot product of the ray and the face unit normal $\hat{n}$ (i.e., $A_{face} \hat{r}\cdot \hat{n}$). Note that regardless of the relative orientation of the faces of this polyhedron, it is always the case that when averaging over all possible orientations of $\hat{n}$ the average area will be a constant times the unprojected area $A_{face}$. This constant is:

\begin{equation} \label{eq:averageArea}
	\frac{1}{4\pi} \int_0^{2\pi} \int_0^{\pi} A_{face} \hat{r}\cdot \hat{n} \sin \theta d\theta d\phi = \frac{1}{2} A_{face} 
\end{equation}

Following that only half of the total polyhedron surface is illuminated by rays at any given angle, and that the cases where the ray is parallel to a surface or an edge of the polyhedron are irrelevant for this analysis because they project to zero area over an infinitesimal angle, we have that the average projected area for the entire polyhedron ($\bar{A}_{tot}$) is:

\begin{equation} \label{eq:averageAreaTot}
	\bar{A}_{tot}=\frac{1}{4} A_{tot}
\end{equation}

This rationale was first presented by \citet{Cauchy1832}. When considering the rays coming from a given face $j$ at all angles, we have exactly the same constant multiplying the unprojected area $A_j$, since only incoming rays are considered (i.e., only a $2\pi$ solid angle is considered). Therefore, the weight related to a given face is:

\begin{equation} \label{eq:averageAreaWeight}
	w_j=\frac{\bar{A}_{j}}{\bar{A}_{tot}}=\frac{A_j}{A_{tot}} 
\end{equation}

\noindent for any convex polyhedron. Note Equation \ref{eq:averageAreaWeight} is true because the $1/4$ constant cancels out. The same immediately follows for 2D polygon elements, although the projection area constant is $1/2\pi$, which also cancels out. 

\begin{figure*} [h]
	\centering
	\includegraphics[width=100mm]{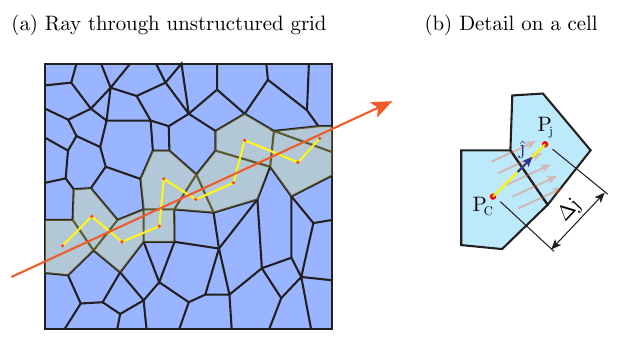}
	\caption{Extension of the omnidirectional method for unstructured grids.}
	\label{fig:unstructured}
\end{figure*}

This means that Equation \ref{eq:FaceCrossing3} can be generalized for arbitrary grid structures made of 3D convex polyhedra or 2D convex polygons, taking a similar form:

\begin{equation} \label{eq:PolyhedralEqn}
	P_C = \sum_j  \frac{\tilde{b}_j A_j}{A_{int}} \bigg\{ P_j - \bigg[\bigg(\frac{\vec{f}(j)+\vec{f}(C)}{2}\bigg)\cdot \hat{j}\bigg] \Delta j  \bigg\}
\end{equation}

\noindent where $j$ is any of the convex polyhedral faces, $\Delta j$ is the distance between cell centroids and $\hat{j}$ is the unit vector in the direction connecting cell centroids. Similarly, $A_{int}$ is the area of the polyhedron faces that are not boundaries. An example of a ray going through a grid made of polygonal elements is presented in Figure \ref{fig:unstructured} (a), with a detail in (b) showing the definitions for Equation \ref{eq:PolyhedralEqn}. Note $\hat{j}$ is not the normal direction of the face, but the direction of the line connecting cell centroids. Similarly, the value of $\vec{f}$ is taken at the mid-point between $C$ and $j$, and not at the face, as it approximates the average value of the source function value through the line connecting $C$ and $j$. This method should enable the treatment of unstructured grids and sparsely measured velocity fields, such as the ones obtained from PTV (Particle Tracking Velocimetry) techniques.

As a simple summary, Equations \ref{eq:FaceCrossing3} and \ref{eq:PolyhedralEqn} both implement an averaging filter, where the pressure at the center cell $P_C$ is the average between the pressures and source terms around all surrounding cells $P_j$, weighed by the area shared between the two cells. The source function is also averaged accordingly. At the boundaries, this same averaging process is performed, though with fewer cells; thereby treating the boundaries in a similar fashion as interior points, to the extent that data is available. Thus, the omnidirectional integration method seems to implement a special boundary condition (i.e., an \textit{``interior point''} boundary condition), which does not appear to be Neumann-like or Dirichlet-like when an arbitrary grid is considered. 

It is not clear that this type of boundary condition is optimal for the arbitrary grid, and it does not seem to implement a finite-volume approximation (divergence theorem) of the Poisson equation, at least to the extent the authors have explored this problem. When the grid is made of cubic elements, as discussed in Section \ref{sec:Poisson}, the Poisson equation and the matrix omnidirectional equation implement the same averaging filter for interior cells. The boundary treatment, however, appears to be superior for the ``omnidirectional'' boundary proposed in Equation \ref{eq:PolyhedralEqn} because it is the same averaging process with fewer points due to the presence of a boundary. In a Neumann boundary, only the forcing normal to the boundary would be considered, and the tangential component is artificially ignored. 
 
\subsection{Unstructured Grid Demonstration} \label{sec:ArbGridsResults}

\begin{figure*} [h]
	\centering
	\includegraphics[width=65mm]{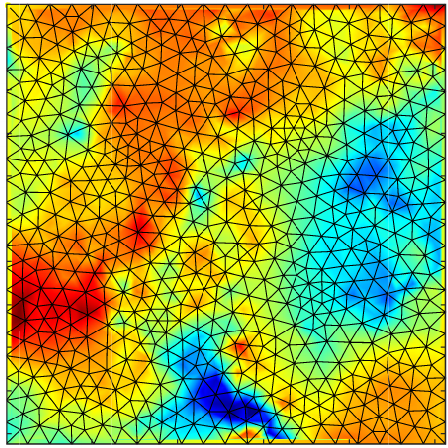}
	\caption{Example of a coarse mesh (1368 triangles) used to demonstrate the OS-MODI method for unstructured grids. The colors in the field represent solved pressure from velocity.}
	\label{fig:Delaunay}
\end{figure*}

To demonstrate the ability of the OS-MODI formulation for unstructured grids delineated in Equation \ref{eq:PolyhedralEqn} to successfully integrate the pressure field without explicitly specifying any boundary conditions, we provide an example with 2D grids consisting of tessellated triangles generated using Matlab's \textit{initmesh} function. Note only the mesh is generated with the Matlab's PDE toolbox, and the solutions are performed with standard matrix inversion. A sample of a coarse grid generated in this manner is provided in Figure \ref{fig:Delaunay}. 

A $512^2$ slice of the JHU turbulence database where the true values for $\vec{u}$ and $P$ are known is prepared as described in Section \ref{sec:OSMODI} to serve as the ground-truth database for the arbitrary grid method. The values required to compute the source function $\vec{f}$ are sampled at the centroids of the triangles generated. We then construct the weight matrix and source vector according to Equation \ref{eq:PolyhedralEqn} in Matlab and solve it using Matlab's default solver \textit{mldivide}.

\begin{figure*} [h]
	\centering
	\hspace{-8mm}\includegraphics[width=170mm]{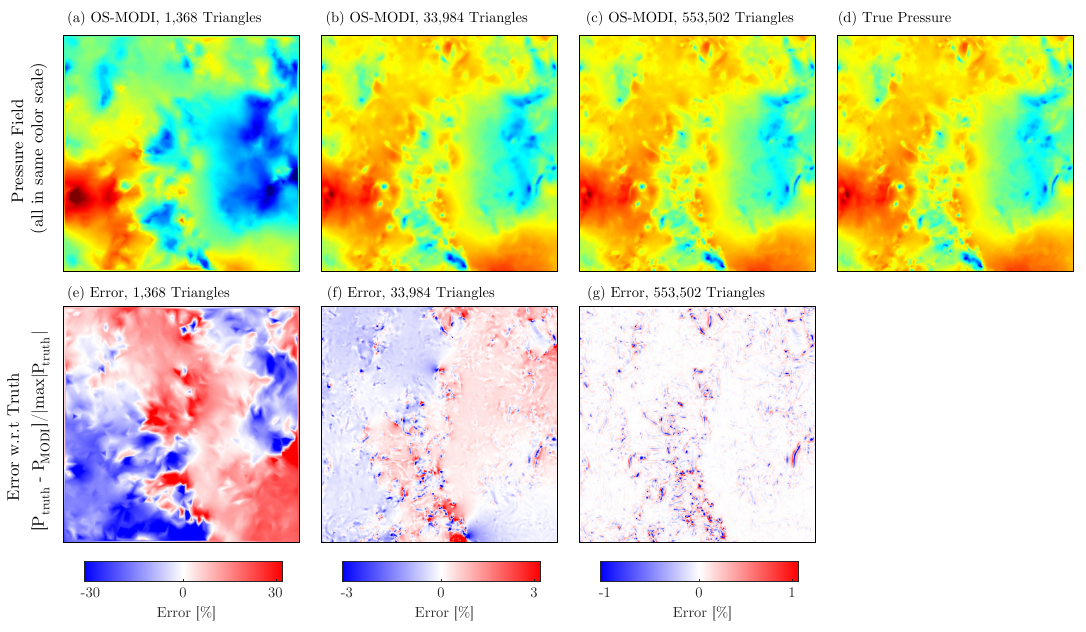}
	\caption{Pressure fields obtained through the unstructured OS-MODI formulation for three mesh refinement levels.}
	\label{fig:ArbSolutions}
\end{figure*}

The pressure field solved through Equation \ref{eq:PolyhedralEqn} is presented for increasingly finer meshes in Figure \ref{fig:ArbSolutions} (a-c) and compared to the true pressure of Figure \ref{fig:ArbSolutions} (d). The corresponding error fields are shown in Figure \ref{fig:ArbSolutions} (e-g). We observe that, even in the coarsest mesh presented, the values of the pressure field are reasonably close to the true pressure and that sufficiently refined sampling leads to very good approximations of the true value of the pressure field with the unstructured OS-MODI formulation. Note the error color scales shown in Figure \ref{fig:ArbSolutions} (e-g) are increasingly tighter as the mesh becomes more refined, demonstrating better approximations of the true pressure with finer meshes.

\begin{figure*} [h]
	\centering
	\hspace{-8mm}\includegraphics[width=170mm]{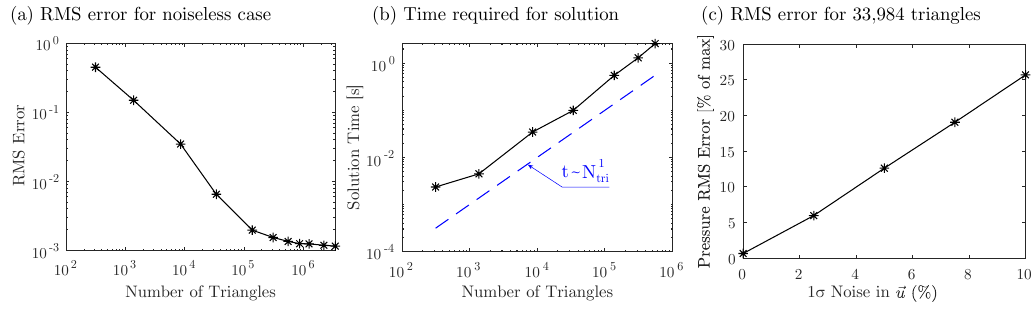}
	\caption{Behavior of error and solution time as a function of unstructured mesh size and error in velocity vectors.}
	\label{fig:ErrorGraphsUnstruct}
\end{figure*}

A quantitative view of the behavior of the RMS error as a function of the mesh size is provided in Figure \ref{fig:ErrorGraphsUnstruct} (a). As expected, the error in the pressure estimates improves with a finer mesh because more information is available for pressure integration. At the larger triangle counts ($>200,000$), the error appears to stabilize to a constant, potentially due to the second-order accuracy of this method. As evidenced in Figure \ref{fig:ErrorGraphsUnstruct} (b), the time required to invert the matrix equation \ref{eq:PolyhedralEqn} increases linearly with the number of triangles in the mesh ($N_{tri}$). This time complexity is very favorable, and is likely to enable the tackling of larger meshes with more efficient code. For this part of the work, however, we did not seek to optimize the computational efficiency of the implementation but only to demonstrate the feasibility of using the OS-MODI method for unstructured grids using Equation \ref{eq:PolyhedralEqn}.

To assess the behavior of error propagation as a function of noise in the underlying velocity vectors, we add noise according to the procedure described in \citet{Zigunov2023mst} and evaluate the RMS pressure error as a function of the maximum absolute pressure value in the field. The results are presented in Figure \ref{fig:ErrorGraphsUnstruct} (c) for the case with 33,984 triangles, demonstrating a linear relationship between the noise in the underlying velocity vectors and the integrated pressure, which is very similar to the trend observed in our past work \citep{Zigunov2023mst} for the I-MODI method; as well as the other ODI methods presented in the literature. 

This demonstrated ability of the OS-MODI method to handle unstructured grids provides the potential to handle pressure calculations on problems where the material derivative is obtained directly from particle tracking (PTV), instead of structured grids such as the more traditional PIV techniques.

\section{Conclusion}

In this study, we extend the recently presented iterative matrix omnidirectional integration solver to be recast such that it solves for pressure in a single shot, by assuming the method converges and ``fast-forwarding'' to the ``infinity'' iteration. This enables the solution of the pressure field using the omnidirectional integration method with an infinite number of rays and eliminating the iterative nature of the method, achieving a computational performance and implementation complexity exactly equal to the pressure-Poisson solver but with the accuracy and robustness to experimental noise of omnidirectional methods, along with the implicit treatment of the boundary conditions that simplifies usage in complex domain shapes with cropped zones. We demonstrate how this class of methods can be extended to arbitrary convex cell meshes, offering an easy extension from gridded PIV data to ungridded data such as found in PTV measurements.

We demonstrate the one-shot method indeed results in the fully-converged solution of the iterative method for the Johns Hopkins turbulence database even in the presence of noise. The resulting method therefore offers identical error characteristics to our previous I-MODI at a fraction of the computational cost. The performance of our GPU implementation enables the solution of very large 3D domains of millions of points in a fraction of a second and up to half a billion grid points in a few minutes, enabling rapid pressure computation from velocity fields in many practical applications. 

Further examination of the equations of the proposed method and comparison to the pressure-Poisson equations reveals significant similarities and a surprisingly simple interpretation for the boundary conditions of the omnidirectional method: it simply treats the boundary cells as close as possible to interior cells, which is further evidenced when examining the generalization of the one-shot matrix method to convex polyhedral grids. It is very likely that the increased robustness of the omnidirectional integration methods observed so far in the literature is related to this new ``interior point'' boundary condition.

Finally, we demonstrate that the extension of the OS-MODI method to convex polyhedral grids is viable and provides accurate estimates of pressure for velocity data sampled within an unstructured grid, potentially enabling the calculation of pressure from PTV measurements with low time complexity.

\section*{Acknowledgements}
	This work was supported by the U.S. Department of Energy through the Los Alamos National Laboratory. Los Alamos National Laboratory is operated by Triad National Security, LLC, for the National Nuclear Security Administration of U.S. Department of Energy, Contract No. 89233218CNA000001

\newcommand{\newblock}{}
\bibliographystyle{apalike}
\bibliography{One-Shot-Pressure}

\begin{thebibliography}{}

\bibitem[Cauchy, 1832]{Cauchy1832}
Cauchy, A.~L. (1832).
\newblock Memoire sur la rectification des courbes et la quadrature des
  surfaces courbes.

\bibitem[Charonko et~al., 2010]{charonko_assessment_2010}
Charonko, J.~J., King, C.~V., Smith, B.~L., and Vlachos, P.~P. (2010).
\newblock Assessment of pressure field calculations from particle image
  velocimetry measurements.
\newblock {\em Measurement Science and Technology}, 21(10):105401.

\bibitem[Li et~al., 2008]{li_turbulence_2008}
Li, Y., Perlman, E., Wan, M., Yang, Y., Meneveau, C., Burns, R., Chen, S.,
  Szalay, A., and Eyink, G. (2008).
\newblock A public turbulence database cluster and applications to study
  lagrangian evolution of velocity increments in turbulence.
\newblock {\em Journal of Turbulence}, 9:N31.

\bibitem[Liu and Katz, 2003]{liu_measurements_2003}
Liu, X. and Katz, J. (2003).
\newblock Measurements of {Pressure} {Distribution} by {Integrating} the
  {Material} {Acceleration}.
\newblock In {\em Fifth {International} {Symposium} on {Cavitation}
  ({CAV2003})}, Osaka, Japan.

\bibitem[Liu and Katz, 2006a]{Liu2006}
Liu, X. and Katz, J. (2006a).
\newblock Instantaneous pressure and material acceleration measurements using a
  four-exposure piv system.
\newblock {\em Experiments in Fluids}, 41(2):227--240.

\bibitem[Liu and Katz, 2006b]{liu_instantaneous_2016}
Liu, X. and Katz, J. (2006b).
\newblock Instantaneous pressure and material acceleration measurements using a
  four-exposure {PIV} system.
\newblock {\em Experiments in Fluids}, 41(2):227--240.

\bibitem[Liu and Moreto, 2020]{liu_error_2020}
Liu, X. and Moreto, J.~R. (2020).
\newblock Error propagation from the {PIV}-based pressure gradient to the
  integrated pressure by the omnidirectional integration method.
\newblock {\em Measurement Science and Technology}, 31(5):055301.

\bibitem[Pan et~al., 2016]{pan_error_2016}
Pan, Z., Whitehead, J., Thomson, S., and Truscott, T. (2016).
\newblock Error propagation dynamics of {PIV}-based pressure field
  calculations: {How} well does the pressure {Poisson} solver perform
  inherently?
\newblock {\em Measurement Science and Technology}, 27(8):084012.

\bibitem[van Oudheusden, 2008]{vanOudheusden2008}
van Oudheusden, B.~W. (2008).
\newblock Principles and application of velocimetry-based planar pressure
  imaging in compressible flows with shocks.
\newblock {\em Experiments in Fluids}, 45(4):657--674.

\bibitem[Wang et~al., 2019]{wang_gpu-based_2019}
Wang, J., Zhang, C., and Katz, J. (2019).
\newblock {GPU}-based, parallel-line, omni-directional integration of measured
  pressure gradient field to obtain the {3D} pressure distribution.
\newblock {\em Experiments in Fluids}, 60(4):58.

\bibitem[Zigunov and Charonko, 2023]{Zigunov2023mst}
Zigunov, F. and Charonko, J.~J. (2023).
\newblock A fast, matrix-based method to perform omnidirectional pressure
  integration {[Under Review, see https://arxiv.org/pdf/2311.16935.pdf]}.
\newblock {\em Measurement Science and Technology}.

\end{thebibliography}

\end{document}